\documentclass[12pt]{article}

\usepackage[margin=1in]{geometry}

\usepackage{setspace}

\usepackage{graphicx}

\usepackage{enumitem}

\usepackage{microtype}
\usepackage{subfigure}
\usepackage{booktabs}
\usepackage{comment}
\usepackage{wrapfig}
\usepackage{arydshln}
\usepackage{enumitem}
\usepackage{multirow}
\usepackage{url}
\usepackage{natbib}

\RequirePackage[
  colorlinks,
  citecolor=blue,
  linkcolor=blue,
  urlcolor=blue,
  pagebackref,
  breaklinks=true
]{hyperref}
\usepackage[hyphenbreaks]{breakurl}
\usepackage{amsmath}
\usepackage{amssymb}
\usepackage{mathtools}
\usepackage{amsfonts}
\usepackage{bm}
\usepackage{bbm}

\usepackage{algorithm}
\usepackage{algpseudocode}
\def\eqref#1{equation~\ref{#1}}
\def\1{\bm{1}}

\DeclareMathAlphabet{\mathsfit}{\encodingdefault}{\sfdefault}{m}{sl}
\SetMathAlphabet{\mathsfit}{bold}{\encodingdefault}{\sfdefault}{bx}{n}

\def\bbE{{\mathbb{E}}}

\usepackage{amsthm}

\theoremstyle{plain}

\newtheorem{theorem}{Theorem}[section]

\newtheorem{proposition}[theorem]{Proposition}

\usepackage[textsize=tiny]{todonotes}
\usepackage{multirow}
\usepackage{wrapfig}
\usepackage{subfigure}

\renewcommand{\eqref}[1]{(\ref{#1})}

\usepackage{xcolor}
\newcount\Comments  % 0 suppresses notes to selves in text
\newcommand{\kibitz}[2]{\ifnum\Comments=1\textcolor{#1}{#2}\fi}

\usepackage{listings}

\usepackage[capitalize,noabbrev]{cleveref}

\allowdisplaybreaks

\title{Sequential Audit Sampling for Finite Populations with Exact and Simulation-based Guarantee}

\author{Masahiro Kato\thanks{Osaka Metropolitan University, Graduate School of Business; RIKEN AIP; Mizuho-DL Financial Technology Co., Ltd.}
\and Kei Nakagawa\thanks{Osaka Metropolitan University, Graduate School of Business.}}

\date{\today}
\begin{document}

\maketitle 

\begin{abstract}
Financial statement auditors use a risk-based approach to evidence collection to obtain reasonable assurance. When an initial sample does not support a conclusion, they may inspect additional items. We develop a sampling procedure by formulating this additional inspection as sequential hypothesis testing for a fixed finite population sampled without replacement. A tolerable deviation rate and a margin below that rate define the composite hypotheses and the intervening indifference region, and lower and upper stopping boundaries control the two probabilities of incorrect conclusions. Under simple random sampling, a hypergeometric recursion computes the boundaries, decision probabilities, and expected sample size exactly. The procedure permits decisions at prespecified times, including fixed intervals. A separate extension imposes a fixed maximum sample size. We also consider one-sided error control and complete examination of selected items. When the state is multidimensional or has too many possible values for direct enumeration, Monte Carlo simulation constructs candidate boundaries. An independent simulation then provides one-sided upper confidence bounds for the two error probabilities at the least-favorable population counts.
\end{abstract}


\section{Introduction}
\label{sec:introduction}

Financial statement auditing plays a central role in protecting investors, lenders, and other stakeholders by providing independent assurance on the reliability of financial reporting. In large organizations, the records underlying account balances and transaction classes can be too numerous to inspect completely. The auditor therefore seeks reasonable assurance rather than absolute assurance and allocates limited audit resources to the areas in which material misstatement is more likely.

Audit sampling is one of the main tools used to obtain that assurance. The auditor applies an audit procedure to less than all items in a population and uses the results to form a conclusion about the population. ISA 530 and related national standards require a reasonable basis for that conclusion and define statistical sampling by random selection and probability-based evaluation of sampling risk \citep{IAASB2025handbookof,ISA530internationalstandard,AUASB2021auditingand,JICPA2023japaneseinstitute}. PCAOB AS 2315 and the \emph{Financial Audit Manual} issued by GAO and CIGIE contain related requirements in the United States \citep{PCAOB2026publiccompany,GAO2026financialauditmanual}.

An initial sample may not provide a sufficient basis for a conclusion. The auditor may then increase the sample size, test an alternative control, or change related substantive procedures. Audit standards recognize each response, but they do not specify stopping boundaries for decisions based on the cumulative sample. Audit-firm policies also differ in their sampling methods, input values, and responses to observed deviations \citep{Christensen2015behindthe,Christensen2015insightsinto}. Differences in confidence levels, expected deviation rates, and tolerable deviation rates can lead to materially different sample sizes. Field evidence further indicates that error rates are low in many control-testing populations and that auditors can have difficulty evaluating sampling risk after deviations are found \citep{Durney2014fielddata,Elder2013auditsampling}.

Auditing research has long described evidence collection as sequential. Normative and descriptive studies treat reliance decisions, evidence search, and evidence evaluation as activities that may change as information is obtained \citep{Kinney1975decisiontheory,Felix1982researchin,Cushing1986comparisonof}. \citet{Ashton1988sequentialbelief} examine sequential belief revision, and \citet{Knechel1990sequentialauditor} examine the auditor's choice between stopping and obtaining additional evidence. These studies show that audit judgments may change as evidence accumulates. They do not give finite-population stopping boundaries with error probabilities fixed before inspection.

Sequential sampling for tests of controls is not new. The AICPA audit sampling guide includes a sequential procedure, and the statistical literature contains sequential tests for hypergeometric finite populations \citep{AICPA2025sequentialsampling,Lai1979sequentialtests,Xiong1995aclass}. \citet{Ignatova2012closedsequential} give exact calculations for closed sequential and multistage procedures with binary outcomes. We do not claim to introduce sequential audit sampling or exact evaluation of an arbitrary set of stopping boundaries.

Our contribution is to construct stopping boundaries from quantities that auditors specify before control testing: the population size, a tolerable deviation rate, a margin below that rate for the favorable region, two limits on incorrect conclusions, and the sample sizes at which an interim decision may be made. In the main analysis, the unfavorable region begins at the first population deviation count above the tolerable count. We use a hypergeometric recursion to construct the boundaries under simple random sampling without replacement and to evaluate the resulting procedure exactly. A monotonicity result extends the error limits from the two population counts that define the favorable and unfavorable regions to the corresponding composite hypotheses.

The two incorrect conclusions have different audit consequences. Selecting the unfavorable hypothesis when the population is favorable corresponds to assessing control risk too high and generally leads to additional work. Selecting the favorable hypothesis when the population deviation rate exceeds the tolerable rate corresponds to assessing control risk too low and can affect audit effectiveness. Separate limits on these probabilities reflect the distinction made in audit standards.

For the basic problem, the hypergeometric recursion computes the boundaries and decision probabilities without simulation error. The supplementary material describes how simulated inspection orders can be used to obtain candidate boundaries when the state needed for exact recursion has too many dimensions or too many possible values. It also gives one-sided binomial confidence bounds based on a separate simulation after the candidate boundaries have been fixed.

Our main objective is to formulate the inspection of additional items as sequential hypothesis testing for a finite population. We control the two probabilities of an incorrect audit-sampling conclusion and calculate the expected sample size. We then examine how the expected sample size changes when decisions are made after each item or at fixed intervals. The numerical example compares the sequential procedures with an exact fixed-sample hypergeometric test under the same two error limits. It also repeats the exact calculation for several population sizes, tolerable deviation rates, and error limits. Further extensions, proofs, and additional calculations appear in the supplementary material.

The remainder of the article proceeds as follows. Section~\ref{sec:setup} defines the finite-population problem. Section~\ref{sec:formulation} formulates the hypotheses and error probabilities. Section~\ref{sec:procedure} presents the exact sequential auditing procedure. Section~\ref{sec:numerical} presents the numerical example and the comparison with a fixed-sample test. Section~\ref{sec:discussion} discusses the conditions under which the procedure is useful and its limitations. Section~\ref{sec:conclusion} concludes.

\section{Problem Setup}
\label{sec:setup}

Suppose that an auditor examines a population of $n$ audit items. Before inspection, the auditor defines the condition that constitutes a deviation. Each item is then classified as a deviation or a nondeviation. In a test of controls, a deviation may be a failure to perform or document the control as specified. Let $X_i\in\{0,1\}$ indicate the classification of item $i$, where $X_i=1$ denotes a deviation.

\subsection{Deviation Rate}

The total number of deviations and the population deviation rate are
\begin{equation}
 m=\sum_{i=1}^{n}X_i
 \quad\text{and}\quad
 p_0=\frac{m}{n}.
 \label{eq:population-rate}
\end{equation}
The values $X_1,\ldots,X_n$ and $m$ are fixed. Probability enters through the order in which the auditor inspects the items.

The auditor selects the inspection order uniformly from all permutations of the population. After relabeling the items by that random order, $X_t$ denotes the result of the $t$th inspection. The cumulative number of deviations and the sample deviation rate are
\begin{equation}
 S_t=\sum_{s=1}^{t}X_s
 \quad\text{and}\quad
 \widehat p_t=\frac{S_t}{t}.
 \label{eq:sample-rate}
\end{equation}
Sampling is without replacement. Conditional on $S_{t-1}=s$, the probability that the next inspected item is a deviation is
\begin{equation}
 \Pr_m(X_t=1\mid S_{t-1}=s)=\frac{m-s}{n-t+1}.
 \label{eq:conditional-probability}
\end{equation}
We write $\Pr_m$ and $\bbE_m$ for probability and expectation under a population with exactly $m$ deviations and a uniformly random inspection order. Equation~\eqref{eq:conditional-probability} is the basis of the exact recursion used below.

\subsection{Tolerable Deviation Rate and Audit Sampling}

Let $r\in(0,1)$ denote the tolerable deviation rate. If all items are inspected, the auditor observes $m$ and compares $p_0$ with $r$ without sampling error. Complete inspection can be costly when $n$ is large, so the auditor may inspect a random sample and infer whether the population supports the intended reliance decision.

The analysis is limited to binary outcomes. Monetary-unit sampling and variables sampling concern a monetary estimand and require different calculations. We discuss their relation to the present procedure in Appendix~\ref{app:existing-studies}, but the binary stopping boundaries below should not be applied to monetary misstatement amounts.

\subsection{Objective}

Our objective is to formulate the inspection of additional items as a sequential auditing procedure. Its stopping and decision rules are fixed before the outcomes are observed. The procedure controls two probabilities of an incorrect conclusion and reports the expected number of inspected items.

The procedure applies when every item not yet inspected is eligible for random selection. If the auditor examines selected high-value or high-risk items in full before random sampling, the deviations found in those items are included in the total deviation count. The results from the selected items are not projected to the remaining population. Appendix~\ref{app:selected-items} gives the calculation for the remaining population.

\section{Formulation as Sequential Testing}
\label{sec:formulation}

Let $\theta_H\geq0$ denote a margin below the tolerable deviation rate. The largest deviation count in the favorable region and the largest count that does not exceed the tolerable rate are
\begin{equation}
 m_H=\left\lfloor n(r-\theta_H)\right\rfloor
 \quad\text{and}\quad
 m_r=\lfloor nr\rfloor.
 \label{eq:hypothesis-counts}
\end{equation}
The count $m_H$ is specified before inspection and defines the favorable region protected by the $\alpha$ limit; it is not an estimate of the unknown population deviation count.
In the main analysis, the unfavorable region begins at
\begin{equation}
 m_K=m_r+1.
 \label{eq:unfavorable-count}
\end{equation}
We assume that $0\leq m_H\leq m_r<n$. The hypotheses are
\begin{equation}
 H:m\leq m_H
 \quad\text{and}\quad
 K:m\geq m_K.
 \label{eq:hypotheses}
\end{equation}
The counts $m_H+1,\ldots,m_r$ form the indifference region. These counts do not exceed the tolerable count. The error limits do not restrict the decision for a population count in this region. Appendix~\ref{app:general-separated} gives the more general separated formulation in which the unfavorable region begins above $m_r+1$.

\subsection{Sequential Sampling Procedure}

A decision time is a number of inspected items at which the cumulative deviation count is compared with the stopping boundaries. At each decision time, the auditor uses $S_t$ to select $H$, select $K$, or inspect additional items. Let $a_t$ and $b_t$ be lower and upper stopping boundaries satisfying $a_t<b_t$. The decision at time $t<n$ is
\begin{equation}
 \delta_t=
 \begin{cases}
 H, & S_t\leq a_t,\\
 K, & S_t\geq b_t,\\
 \text{continue}, & a_t<S_t<b_t.
 \end{cases}
 \label{eq:interim-decision}
\end{equation}
The stopping time is the first decision time at which the result is $H$ or $K$. If no earlier decision is reached, the auditor inspects all $n$ items. At complete inspection, the procedure selects $H$ when $m\leq m_r$ and selects $K$ when $m>m_r$. Let $\delta\in\{H,K\}$ denote the final decision, whether it is reached before or at complete inspection.

The corresponding boundaries for the sample deviation rate are $a_t/t$ and $b_t/t$. We use cumulative counts in the computation because their support is the finite set $\{0,1,\ldots,t\}$.

\subsection{Control of the Error Probabilities}

Two incorrect conclusions are relevant. Selecting $K$ when $H$ holds corresponds to assessing control risk too high. Selecting $H$ when $K$ holds corresponds to assessing control risk too low. Let $\alpha,\beta\in(0,1/2)$ denote the respective limits. We require
\begin{equation}
 \sup_{m\leq m_H}\Pr_m(\delta=K)\leq\alpha
 \quad\text{and}\quad
 \sup_{m\geq m_K}\Pr_m(\delta=H)\leq\beta.
 \label{eq:error-limits}
\end{equation}
Because $m_K=m_r+1$, the second inequality applies to every finite population whose deviation rate exceeds the tolerable deviation rate. The values $\alpha$ and $\beta$ limit the probabilities of the two incorrect sampling conclusions produced by this procedure; they do not represent the auditor's overall audit-risk assessment. The value of $\beta$ may be smaller than $\alpha$ when assessing control risk too low has the more serious audit consequence.

For a fixed set of boundaries, the event of selecting $K$ is monotone in the population deviation count, while the event of selecting $H$ is monotone in the opposite direction. The monotonicity allows the two composite error limits to be checked at $m_H$ and $m_K$.

\begin{theorem}[Least-favorable counts]
\label{thm:least-favorable}
For any stopping boundaries satisfying $a_t<b_t$, $\Pr_m(\delta=K)$ is nondecreasing in $m$, and $\Pr_m(\delta=H)$ is nonincreasing in $m$. The two inequalities in \eqref{eq:error-limits} therefore hold when
\begin{equation}
 \Pr_{m_H}(\delta=K)\leq\alpha
 \quad\text{and}\quad
 \Pr_{m_K}(\delta=H)\leq\beta.
 \label{eq:least-favorable-errors}
\end{equation}
\end{theorem}

Appendix~\ref{app:proofs} gives the proof. The theorem holds for every finite $n$ and uses only sampling without replacement from a fixed population.

\section{Sequential Auditing Procedure}
\label{sec:procedure}

We next compute stopping boundaries that satisfy \eqref{eq:least-favorable-errors}. Exact computation is available because the state at time $t$ is the cumulative deviation count $S_t$.

\subsection{Exact Computation of the Boundaries}

For a population containing $m$ deviations, let
\begin{equation}
 q_{t-1}^{(m)}(s)=\Pr_m\left(S_{t-1}=s,\ a_j<S_j<b_j\text{ for all decision times }j<t\right)
 \label{eq:continuation-mass}
\end{equation}
denote the probability of reaching state $s$ immediately before the $t$th inspection without an earlier decision. We set $q_0^{(m)}(0)=1$.

The probability of reaching state $s$ at time $t$ before applying the boundaries is
\begin{align}
 g_t^{(m)}(s)
 ={}&q_{t-1}^{(m)}(s)\frac{n-m-(t-1-s)}{n-t+1}
 \nonumber\\
 &+q_{t-1}^{(m)}(s-1)\frac{m-s+1}{n-t+1},
 \label{eq:recursion}
\end{align}
where a term is zero when its state is outside the feasible range. The first term corresponds to a nondeviation at time $t$, and the second term corresponds to a deviation.

Suppose that the boundaries before time $t$ have been fixed. For a candidate upper boundary $b$, the probability of first selecting $K$ at time $t$ under $m_H$ is
\begin{equation}
 A_t(b)=\sum_{s=b}^{t}g_t^{(m_H)}(s).
 \label{eq:upper-increment}
\end{equation}
For a candidate lower boundary $a$, the probability of first selecting $H$ at time $t$ under $m_K$ is
\begin{equation}
 B_t(a)=\sum_{s=0}^{a}g_t^{(m_K)}(s).
 \label{eq:lower-increment}
\end{equation}
The first-stopping events at different times are disjoint. We therefore choose the smallest feasible $b_t$ and the largest feasible $a_t$ such that
\begin{equation}
 \sum_{j<t}A_j(b_j)+A_t(b_t)\leq\alpha
 \quad\text{and}\quad
 \sum_{j<t}B_j(a_j)+B_t(a_t)\leq\beta.
 \label{eq:recursive-boundaries}
\end{equation}
If these two boundary choices give $a_t\geq b_t$, we replace them by adjacent boundaries $(c,c+1)$. We choose the feasible value of $c$ closest to $tm_r/n$, using the smaller integer when two values are equally close. Lowering the lower boundary or raising the upper boundary cannot increase either error probability.

After choosing $a_t$ and $b_t$, the continuation probabilities are
\begin{equation}
 q_t^{(m)}(s)=g_t^{(m)}(s)\bm{1}(a_t<s<b_t).
 \label{eq:update-continuation}
\end{equation}
Here, $\bm{1}(\cdot)$ is the indicator function. The recursion continues through time $n-1$. Complete inspection at time $n$ gives the terminal decision described in Section~\ref{sec:formulation}.

\begin{proposition}[Exact error control]
\label{prop:exact-control}
The recursive boundary construction in \eqref{eq:recursive-boundaries} satisfies the error limits in \eqref{eq:error-limits}.
\end{proposition}

The proof follows because the cumulative first-stopping probabilities are bounded at $m_H$ and $m_K$, and Theorem~\ref{thm:least-favorable} extends these bounds to both composite regions.

Algorithm~\ref{alg:exact} summarizes the calculation.

\begin{algorithm}[t]
\caption{Exact computation of the stopping boundaries}
\label{alg:exact}
\begin{algorithmic}[1]
\Require $n$, $m_H$, $m_K$, $m_r$, $\alpha$, $\beta$, and the decision times
\State Set the continuation probability at time zero equal to one at state zero for $m_H$ and $m_K$
\For{$t=1,\ldots,n-1$}
  \State Use \eqref{eq:recursion} to compute the probabilities before a decision under $m_H$ and $m_K$
  \If{$t$ is a decision time}
    \State Choose $b_t$ and $a_t$ according to \eqref{eq:recursive-boundaries}
  \Else
    \State Set $a_t=-1$ and $b_t=t+1$
  \EndIf
  \State Retain only the states satisfying $a_t<S_t<b_t$
\EndFor
\State At time $n$, select $H$ when $m\leq m_r$ and select $K$ otherwise
\end{algorithmic}
\end{algorithm}

For the present one-dimensional state space, the recursion computes the boundaries without simulation error. Appendix~\ref{app:monte-carlo} describes a Monte Carlo approximation for settings in which the state cannot be represented by one cumulative deviation count or contains too many possible values for direct enumeration.

\section{Numerical Example}
\label{sec:numerical}

\subsection{Basic Setting}

We set
\begin{equation}
 n=100,\qquad r=0.20,\qquad \theta_H=0.05,\qquad \alpha=\beta=0.05.
 \label{eq:basic-setting}
\end{equation}
The favorable region ends at $m_H=15$, the tolerable count is $m_r=20$, and the unfavorable region begins at $m_K=21$. Thus, the second error limit applies to every population deviation count above the tolerable count. The setting is illustrative and is not intended to reproduce a particular audit-firm policy.

\begin{figure}[htbp]
\centering
\includegraphics[width=0.88\linewidth]{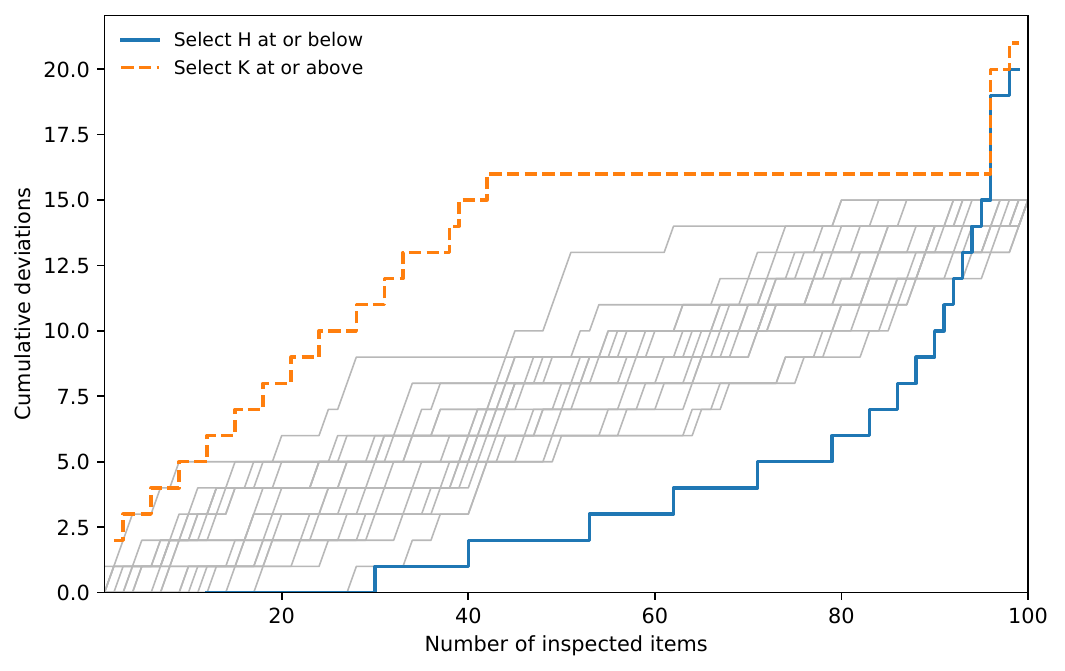}
\caption{Stopping boundaries and inspection paths for a population with 15 deviations.}
\label{fig:boundaries-h}
\end{figure}

\begin{figure}[htbp]
\centering
\includegraphics[width=0.88\linewidth]{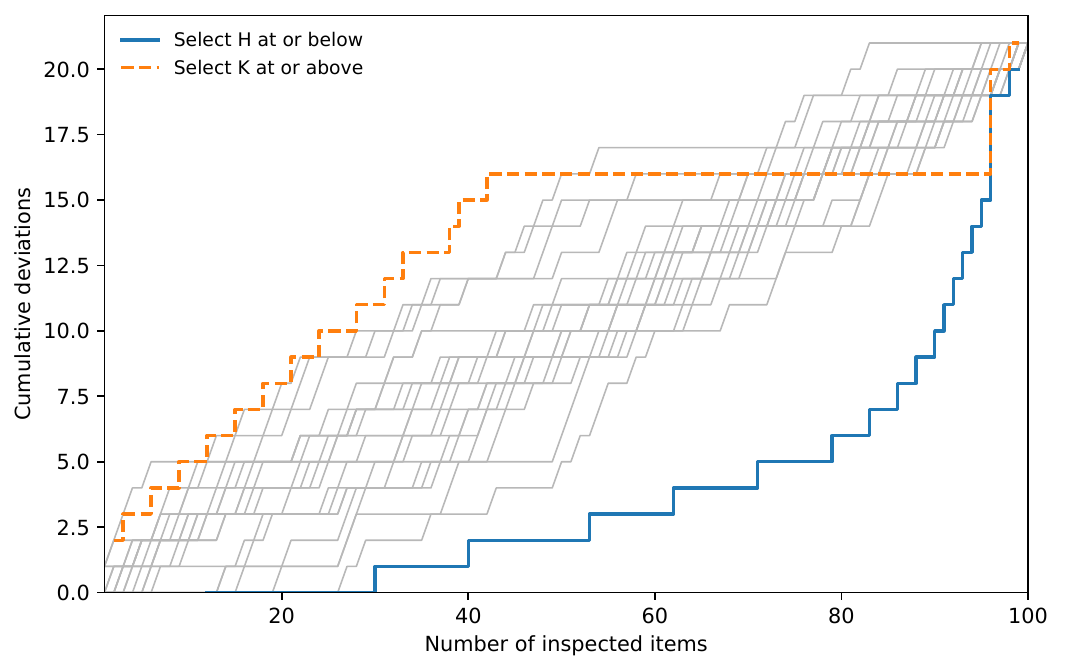}
\caption{Stopping boundaries and inspection paths for a population with 21 deviations.}
\label{fig:boundaries-k}
\end{figure}

Figures~\ref{fig:boundaries-h} and \ref{fig:boundaries-k} show the exact stopping boundaries and 20 randomly generated inspection paths at $m_H$ and $m_K$. Each path is a permutation of a fixed population, so all paths in Figure~\ref{fig:boundaries-h} end at 15 deviations and all paths in Figure~\ref{fig:boundaries-k} end at 21 deviations.

\subsection{Decision Probabilities and Expected Sample Size}

\begin{figure}[htbp]
\centering
\includegraphics[width=0.76\linewidth]{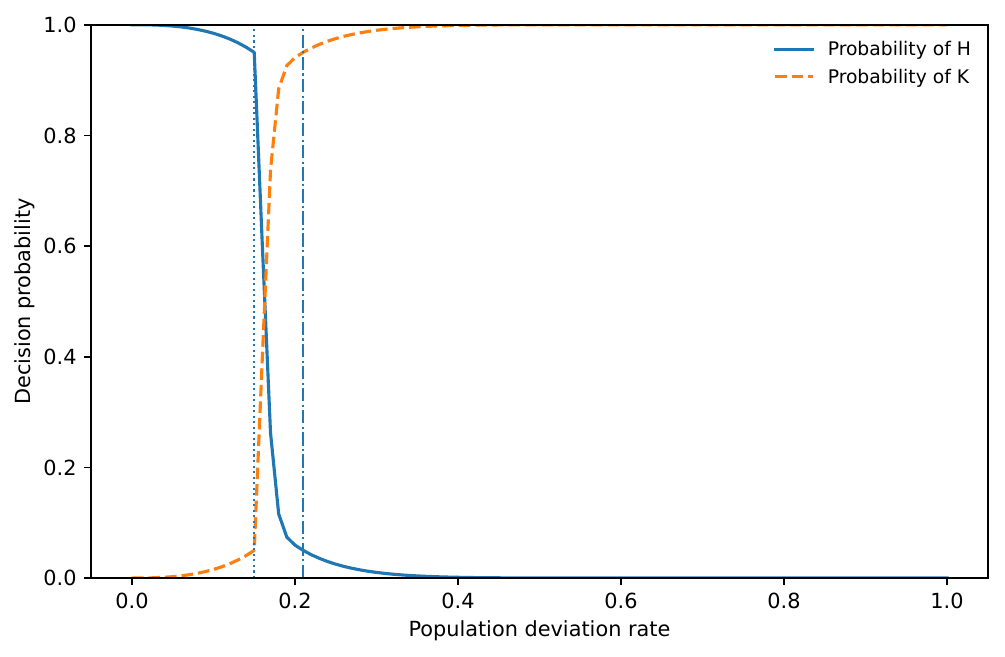}
\caption{Exact decision probabilities over all population deviation counts.}
\label{fig:decision-probabilities}
\end{figure}

\begin{figure}[htbp]
\centering
\includegraphics[width=0.76\linewidth]{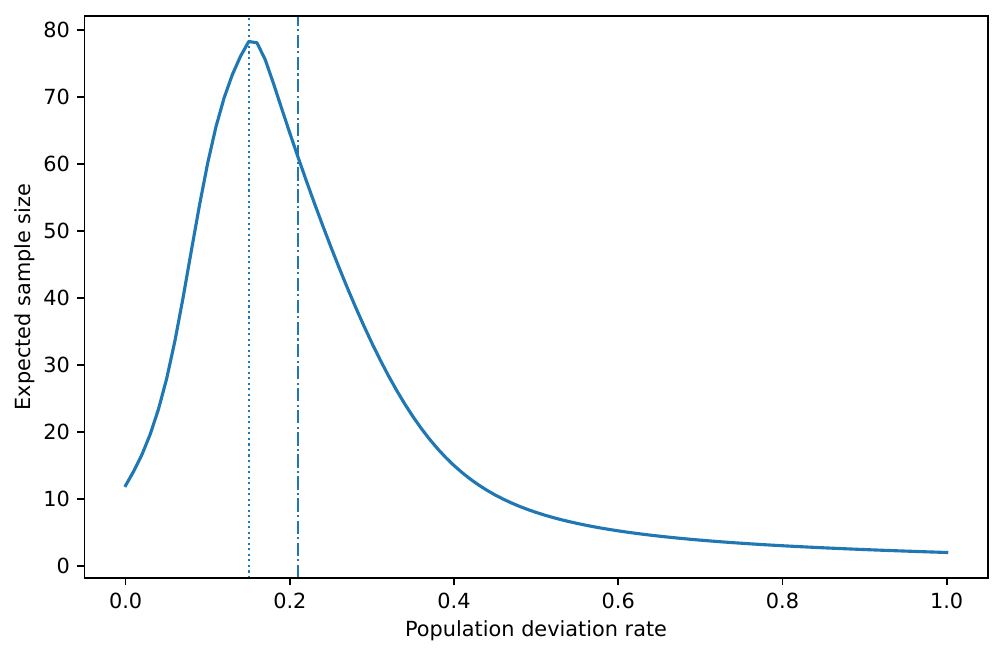}
\caption{Expected sample size over all population deviation counts.}
\label{fig:expected-sample}
\end{figure}

Figure~\ref{fig:decision-probabilities} reports the exact probabilities of selecting $H$ and $K$ over the complete deviation-count grid. The first error limit applies for $m\leq15$, and the second applies for every $m\geq21$. The largest probability of selecting $K$ over $m\leq15$ occurs at $m=15$, and the largest probability of selecting $H$ over $m\geq21$ occurs at $m=21$. Both probabilities equal 0.05 up to numerical precision. Figure~\ref{fig:expected-sample} reports the expected sample size.

The expected sample size is 27.96 when $m=5$, 78.28 when $m=15$, 64.61 when $m=20$, 61.02 when $m=21$, and 33.26 when $m=30$. The procedure tends to stop early when the population deviation count is well below $m_H$ or well above $m_K$. Counts in or near the indifference region require more inspection.

\subsection{Comparison with an Exact Fixed-Sample Test}

The smallest fixed-sample hypergeometric test satisfying both error limits uses 82 items and selects $H$ when the sample contains at most 14 deviations. Its probability of selecting $K$ at $m=15$ is 0.03934, and its probability of selecting $H$ at $m=21$ is 0.04614. Table~\ref{tab:fixed-sample-comparison} compares expected sample sizes at selected population counts.

\begin{table}[htbp]
\centering
\caption{Expected sample size in the basic setting}
\label{tab:fixed-sample-comparison}
\begin{tabular}{lrrrrr}
\toprule
Procedure & $m=5$ & $m=15$ & $m=20$ & $m=21$ & $m=30$ \\
\midrule
Sequential, decision after each item & 27.96 & 78.28 & 64.61 & 61.02 & 33.26 \\
Exact fixed-sample test & 82.00 & 82.00 & 82.00 & 82.00 & 82.00 \\
\bottomrule
\end{tabular}
\end{table}

In the basic setting, the sequential procedure has an expected sample size below 82 for every population deviation count $m=0,\ldots,100$. The largest value is 78.28 at $m=15$. Table~\ref{tab:fixed-sample-comparison} reports selected counts, and Figure~\ref{fig:expected-sample} gives the complete grid. This result is specific to the basic setting. Table~\ref{tab:audit-inputs} shows that an exact fixed-sample test can require fewer items near the decision boundary when the population is larger.

\subsection{Timing of Interim Decisions}

We next permit decisions after every item and after every 5, 10, or 20 items. Each procedure is computed separately under the same error limits. Table~\ref{tab:decision-times} reports the expected sample size and the expected number of decision times reached at $m_H$ and $m_K$.

\begin{table}[htbp]
\centering
\caption{Timing of interim decisions in the basic setting}
\label{tab:decision-times}
\begin{tabular}{lrrrr}
\toprule
Decision interval & $\bbE_{15}[\tau]$ & $\bbE_{21}[\tau]$ & $\bbE_{15}[D]$ & $\bbE_{21}[D]$ \\
\midrule
1 item & 78.28 & 61.02 & 78.28 & 61.02 \\
5 items & 73.62 & 56.69 & 14.72 & 11.34 \\
10 items & 75.55 & 61.14 & 6.99 & 6.11 \\
20 items & 70.84 & 64.32 & 3.24 & 3.06 \\
\bottomrule
\end{tabular}
\begin{flushleft}
\footnotesize Notes: $D$ denotes the number of interim decision times reached. Each row uses boundaries computed for the stated interval.
\end{flushleft}
\end{table}

\begin{figure}[htbp]
\centering
\includegraphics[width=0.78\linewidth]{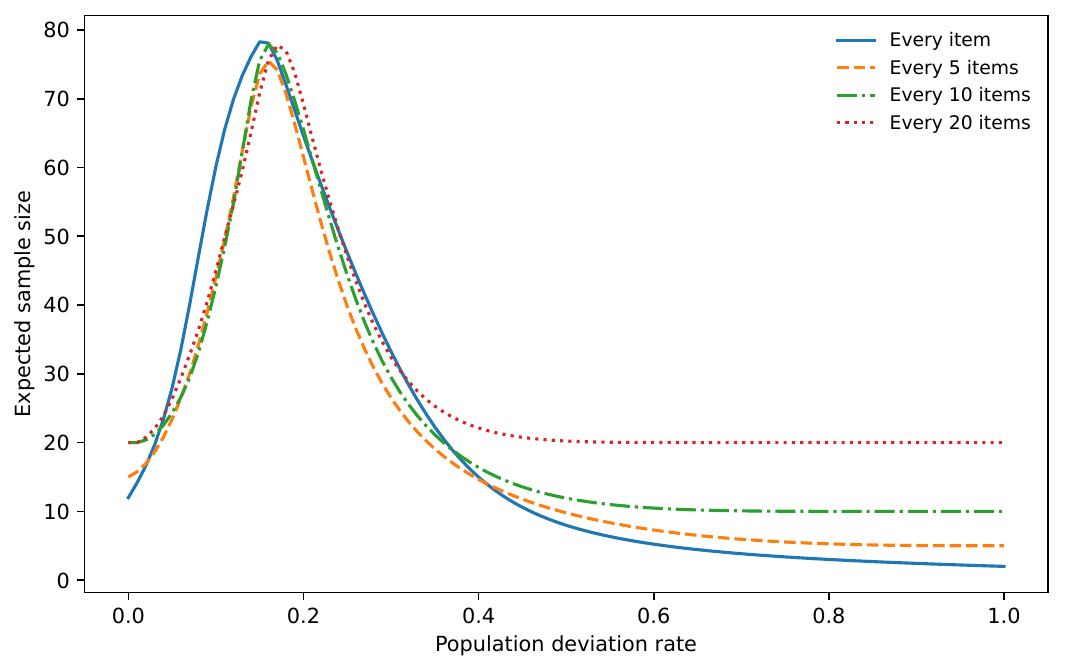}
\caption{Expected sample size for different decision intervals.}
\label{fig:decision-times-sample}
\end{figure}

\begin{figure}[htbp]
\centering
\includegraphics[width=0.78\linewidth]{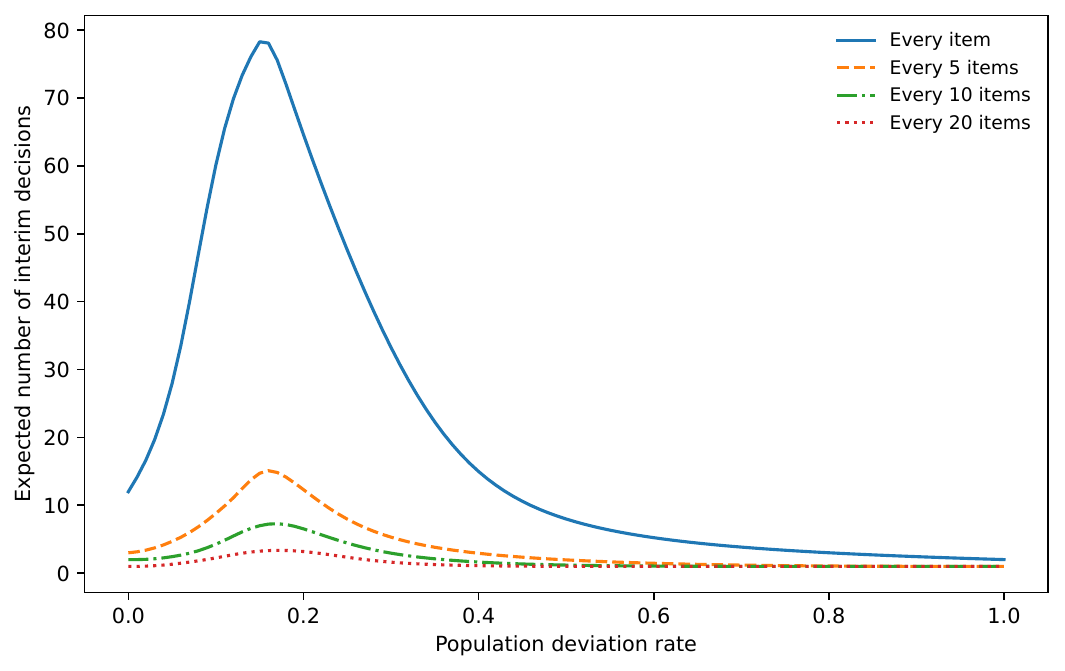}
\caption{Expected number of interim decision times reached.}
\label{fig:decision-times-count}
\end{figure}

Under the boundary construction used here, decisions every five items reduce the expected sample size at both boundary counts relative to decisions after every item. Decisions every 20 items reduce the expected sample size at $m=15$ but increase it at $m=21$. The number of interim decision times reached falls sharply as the interval increases. Among the four sets of decision times examined here, no set has the smallest expected sample size for every population deviation count.

\subsection{Population Size and the Tolerable Deviation Rate}

Audit standards identify the tolerable deviation rate, the expected deviation rate, and the sampling risk that the auditor is willing to accept as factors that affect sample size \citep{JICPA2023japaneseinstitute}. We therefore repeat the exact calculations for population sizes of 100, 500, and 1,000 and tolerable deviation rates of 0.05, 0.10, and 0.20. For comparability across settings, we set $m_H=\lfloor nr/2\rfloor$ and set both error limits to 0.05. This choice is illustrative and is not an estimate of an audit firm's expected deviation rate.

\begin{table}[htbp]
\centering
\small
\caption{Expected sample size for different population sizes and tolerable deviation rates}
\label{tab:audit-inputs}
\begin{tabular}{rrr rr rr}
\toprule
& & & \multicolumn{2}{c}{Decision after each item} & \multicolumn{2}{c}{Decision every 5 items} \\
$n$ & $r$ & Fixed sample & $\bbE_{m_H}[\tau]$ & $\bbE_{m_K}[\tau]$ & $\bbE_{m_H}[\tau]$ & $\bbE_{m_K}[\tau]$ \\
\midrule
100  & 0.05 & 73  & 63.12  & 36.30  & 62.62  & 37.91 \\
100  & 0.10 & 72  & 62.60  & 43.60  & 57.62  & 38.50 \\
100  & 0.20 & 57  & 54.63  & 37.95  & 41.66  & 32.02 \\
500  & 0.05 & 261 & 307.02 & 203.16 & 272.14 & 177.12 \\
500  & 0.10 & 185 & 281.49 & 213.78 & 222.66 & 179.79 \\
500  & 0.20 & 107 & 213.48 & 169.21 & 166.96 & 99.66 \\
1000 & 0.05 & 372 & 628.93 & 426.99 & 592.92 & 372.13 \\
1000 & 0.10 & 227 & 503.43 & 417.90 & 418.01 & 345.13 \\
1000 & 0.20 & 121 & 306.18 & 271.21 & 253.26 & 196.34 \\
\bottomrule
\end{tabular}
\begin{flushleft}
\footnotesize Notes: The favorable count is $m_H=\lfloor nr/2\rfloor$, and the unfavorable count is $m_K=\lfloor nr\rfloor+1$. Both error limits are 0.05. The fixed-sample column gives the exact fixed sample size. The row with $n=100$ and $r=0.20$ uses $m_H=10$ and therefore differs from the basic setting, which uses $m_H=15$.
\end{flushleft}
\end{table}

The sequential procedures do not have smaller expected sample sizes in every setting. At $n=100$, the sequential procedures use fewer items than the fixed-sample test at both $m_H$ and $m_K$. At $n=500$ or 1,000, the exact fixed-sample test can require fewer items near these counts, especially at $m_H$. By contrast, when $m=0$, decisions after each item give expected sample sizes between 12 and 56 across the nine settings, whereas the exact fixed sample sizes range from 57 to 372. The accompanying data repeat the calculation with both error limits set to 0.10. Increasing both error limits reduces the sample sizes, but the ordering depends on the population count and the times at which decisions are allowed.

\section{Discussion}
\label{sec:discussion}

The main result concerns sampling risk under simple random sampling without replacement. The procedure fixes the stopping boundaries before inspection and controls the two incorrect conclusions over the favorable and unfavorable regions. Because $m_K=m_r+1$ in the main analysis, the probability of selecting $H$ is controlled for every population deviation rate above the tolerable rate.

The numerical example also shows the limits of the procedure. Sequential inspection is most useful when the deviation count is sufficiently far from the decision boundary. Near that boundary, more evidence is needed, and the expected sample size can approach complete inspection. Table~\ref{tab:audit-inputs} shows that a fixed-sample test can require fewer items near the decision boundary when the population is large. The timing of interim decisions also changes the boundaries because each set of decision times uses the two cumulative error limits differently. The auditor should therefore specify the decision times before inspection rather than add decision times after observing the sample results.

Before inspection, the auditor must define the population and the condition that constitutes a deviation, set the tolerable deviation rate and the two error limits, choose the decision times, and record the resulting boundaries in the audit documentation. At each decision time, the cumulative deviation count is compared with those boundaries. Selection of $K$ means that the accumulated sample does not support the intended reliance conclusion under the stated error limits. It does not determine the auditor's subsequent response, which may include increasing the sample size, testing another control, or changing related substantive procedures.

The boundary construction does not minimize expected sample size over all possible sequential procedures. Its purpose is to translate the population size, the endpoint $m_H$ of the favorable region, the tolerable deviation rate, the decision times, and the two error limits into stopping boundaries whose risks can be calculated exactly. Prior control-testing results, changes in the control, and the expected deviation rate used in planning may inform the choice of $m_H$, but $m_H$ is specified before inspection and is not an estimate of the population deviation count.

The procedure does not address nonsampling risk. Errors in defining the population, classifying deviations, applying the audit procedure, or interpreting the evidence can lead to an incorrect audit conclusion. The auditor must investigate the nature and cause of observed deviations and determine how they affect the reliance decision and other audit procedures.

The supplementary material considers one-sided error control, a fixed maximum sample size, complete examination of selected items, and a Monte Carlo calculation. It also gives an illustration using CMS CERT claims. Those calculations extend or illustrate the main procedure; they are not evidence from completed audit engagements.

\section{Conclusion}
\label{sec:conclusion}

We formulate audit sampling with sequentially inspected items as a sequential test for a finite population sampled without replacement. The favorable region ends below the tolerable count, and the unfavorable region begins at the first count above it. The procedure controls the risks of assessing control risk too high and too low while permitting an early conclusion. In the basic setting, a hypergeometric recursion computes the boundaries, decision probabilities, and expected sample size exactly. Sequential inspection can reduce expected sample size when the deviation count is far from the decision boundary, whereas an exact fixed-sample test can require fewer items near that boundary in larger populations. The effect of fixed decision intervals also depends on the population deviation count.

The procedure controls sampling risk under the stated random-selection rule. It does not replace investigation of the nature and cause of deviations or the auditor's assessment of their effect on other audit procedures. Within the stated sampling model, the stopping boundaries provide a statistical basis for deciding when to inspect additional items and when the accumulated sample supports a conclusion. The auditor must separately decide how an unfavorable conclusion affects further tests of controls and substantive procedures.

\bibliography{arXiv.bbl}

\bibliographystyle{plainnat}

\clearpage
\appendix

\ifnum\value{section}>6\relax
\appendix
\fi

\section{Institutional Background and Existing Studies}
\label{app:existing-studies}

\subsection{Audit Standards and Audit-Firm Practice}

Audit standards define audit sampling as the application of audit procedures to less than all items in a population so that the auditor can form a reasonable basis for a population conclusion \citep{ISA530internationalstandard,JICPA2023japaneseinstitute,PCAOB2026publiccompany}. Statistical sampling requires random selection and probability-based evaluation of sampling risk. When sampling does not provide a reasonable basis, the auditor may increase the sample size, test an alternative control, or change substantive procedures.

The audit-firm evidence in \citet{Christensen2015behindthe} and \citet{Christensen2015insightsinto} shows variation in sampling methods, confidence levels, deviation-rate inputs, and responses to detected deviations. Some firms permit additional control testing, whereas others more commonly change substantive procedures. The field data in \citet{Durney2014fielddata} cover 160 sampling applications and show low error rates in many post-SOX populations. Variation in firm guidance and the low error rates in those applications motivate examination of a procedure that may stop early when deviations are few while keeping sampling risk below a stated limit.

\subsection{Finite-Population Sequential Testing}

\citet{Lai1979sequentialtests} studies sequential tests for hypergeometric distributions and finite populations. \citet{Xiong1995aclass} develops sequential conditional probability-ratio tests whose maximum sample size is no larger than that of a fixed-sample test with a similar power function. \citet{Ignatova2012closedsequential} give exact calculations for closed sequential and multistage procedures with binary outcomes, with or without replacement. These studies provide the statistical foundation for sequential finite-population analysis. The present article starts from a tolerable deviation rate, composite hypotheses, separate error limits, and specified decision times, and uses them to produce the complete stopping rule. The AICPA guide uses different inputs and a different terminal rule. The numerical comparison in the article therefore uses an exact fixed-sample test constructed under the same finite-population error limits.

\begin{table}[H]
\centering
\small
\caption{Relation to selected existing studies}
\label{tab:existing-studies}
\begingroup\sloppy\begin{tabular}{@{}p{0.27\linewidth}p{0.32\linewidth}p{0.35\linewidth}@{}}
\toprule
Study & Setting and result & Relation to this article \\
\midrule
AICPA audit sampling guide & Includes a sequential procedure for tests of controls. & We give a finite-population calculation for boundaries constructed from the audit inputs used in the present analysis. \\
\citet{Lai1979sequentialtests} and \citet{Xiong1995aclass} & Develop sequential tests for hypergeometric or finite populations. & We use a tolerable deviation rate to define the protected population counts and allow the two incorrect conclusions to have different limits. \\
\citet{Ignatova2012closedsequential} & Gives exact calculations for specified closed sequential and multistage procedures. & We construct the boundaries recursively from the two error limits and the chosen decision times. \\
\citet{WaudbySmith2021rilasrisk} & Audits a fixed population of ballots and proceeds to a complete hand count if necessary. & Both settings use sampling without replacement and an exact terminal determination; our hypotheses concern a tolerable deviation rate in tests of controls. \\
\citet{Shekhar2023risklimiting} & Studies weighted financial auditing with monetary outcomes and unequal selection probabilities. & Our main analysis concerns binary deviations and simple random sampling without replacement. \\
\bottomrule
\end{tabular}\endgroup
\end{table}

\subsection{Sequential Auditing in Other Applications}

Recent research applies sequential auditing to elections, financial transactions, fairness, and privacy. \citet{WaudbySmith2021rilasrisk} studies risk-limiting election audits for a fixed population of ballots sampled without replacement. The election-audit setting is close to ours because a complete hand count determines the election outcome if the audit does not stop early. \citet{Shekhar2023risklimiting} studies weighted financial auditing under sampling without replacement and uses information available before inspection to improve sampling efficiency. That work is the closest reference for extensions to monetary outcomes and unequal inclusion probabilities.

\citet{Chugg2023auditingfairness} studies fairness auditing under probabilistic data collection that can change over time. \citet{Gonzalez2025sequentiallyauditing} studies repeated black-box experiments for detecting violations of differential privacy. Together, these papers show how sequential auditing has been developed for several audit targets and data-collection settings. The present article concerns binary control deviations in a fixed finite population sampled without replacement.

\section{Extensions}
\label{app:extensions}

\subsection{General Separated Hypotheses}
\label{app:general-separated}

The main analysis begins the unfavorable region at the first count above the tolerable count. A more general separated formulation can use two nonnegative margins, $\theta_H$ and $\theta_K$, and define
\begin{equation}
 m_H=\left\lfloor n(r-\theta_H)\right\rfloor
 \quad\text{and}\quad
 m_K=\left\lceil n(r+\theta_K)\right\rceil.
 \label{eq:general-separated-counts}
\end{equation}
The hypotheses are $H:m\leq m_H$ and $K:m\geq m_K$. When $m_K>m_r+1$, the second error limit applies only to $m\geq m_K$. It does not restrict the probability of selecting $H$ for the above-tolerable counts $m_r+1,\ldots,m_K-1$. The main analysis avoids that limitation by setting $m_K=m_r+1$.

\subsection{One-Sided Procedure}

In some audits, the principal concern is selecting $H$ when the population deviation rate exceeds the tolerable rate. A one-sided procedure permits an early selection of $H$ and otherwise continues to complete inspection. The smallest unacceptable count is $m_r+1$. Lower boundaries are chosen so that the cumulative probability of selecting $H$ at this count does not exceed $\beta$. Monotonicity then gives
\begin{equation}
 \sup_{m\geq m_r+1}\Pr_m(\text{early selection of }H)\leq\beta.
 \label{eq:one-sided-error}
\end{equation}
If the procedure does not stop early, complete inspection gives the exact conclusion.

\subsection{Procedure with a Fixed Maximum Sample Size}
\label{app:fixed-maximum}

The main procedure continues to complete inspection if it does not stop early. The auditor can instead set a maximum sample size $T<n$ and require a terminal decision at $T$. The early boundaries and the terminal cutoff must be constructed jointly because both contribute to the two error probabilities.

Let $\alpha_I<\alpha$ and $\beta_I<\beta$ be the portions used for decisions before $T$. We first construct the interim boundaries through $T-1$ under these limits. Let $q_T^{(m_H)}$ and $q_T^{(m_K)}$ denote the probability distributions that reach time $T$ without an earlier decision. A terminal cutoff $c_T$ selects $H$ when $S_T\leq c_T$ and selects $K$ otherwise. The cutoff must satisfy
\begin{equation}
 \sum_{t<T}A_t(b_t)+\Pr_{m_H}(S_T>c_T,\tau\geq T)\leq\alpha
 \label{eq:fixed-maximum-alpha}
\end{equation}
and
\begin{equation}
 \sum_{t<T}B_t(a_t)+\Pr_{m_K}(S_T\leq c_T,\tau\geq T)\leq\beta.
 \label{eq:fixed-maximum-beta}
\end{equation}
For each candidate maximum sample size $T$, the code evaluates the integer terminal cutoffs and retains the smallest $T$ for which both total error limits are satisfied.

\subsection{Complete Examination of Selected Items}
\label{app:selected-items}

An auditor may examine selected high-value or high-risk items in full before drawing a random sample from the remaining population. Let $n_S$ denote the number of selected items and let $m_S$ denote the deviations found among them. The residual population has size $n_R=n-n_S$. The corresponding counts for the residual population are
\begin{equation}
 m_{H,R}=m_H-m_S
 \quad\text{and}\quad
 m_{K,R}=m_K-m_S.
 \label{eq:residual-counts}
\end{equation}
The terminal tolerable count for the residual population is $m_r-m_S$. If $m_{H,R}<0$, the selected items alone rule out the favorable region. If $m_{K,R}>n_R$, the remaining population cannot make the total count reach the unfavorable region. Otherwise, the exact recursion is applied to the residual population. The observations from the selected items are not projected to the remaining items.

\subsection{Monte Carlo Computation}
\label{app:monte-carlo}

When the state cannot be represented by one cumulative deviation count, direct enumeration may require too many states. Examples include sampling from many strata, sampling with unequal selection probabilities, and outcomes that are not binary. In these settings, simulated inspection orders can be used to obtain candidate boundaries.

For the upper boundary, construct a finite population with exactly $m_H$ deviations and generate $M$ independent random permutations. For the lower boundary, use a finite population with exactly $m_K$ deviations. At each candidate boundary, estimate $A_t(b)$ and $B_t(a)$ by the proportion of all $M$ permutations that first cross the corresponding boundary at time $t$. Each estimate uses $M$ as the denominator, including permutations that stopped earlier. The permutations used to construct the boundaries are not reused for the subsequent confidence-bound calculation.

Because the number of simulated permutations is finite, the selected boundaries can have true error probabilities above the targets. We therefore use internal limits $\alpha_0<\alpha$ and $\beta_0<\beta$ when computing candidate boundaries. After fixing those boundaries, we generate an independent set of random permutations at each protected count. If $C_K$ errors occur among $M_H^{\mathrm{sim}}$ permutations generated at $m_H$, and $C_H$ errors occur among $M_K^{\mathrm{sim}}$ permutations generated at $m_K$, let $U_K$ and $U_H$ be one-sided Clopper--Pearson upper confidence bounds \citep{Clopper1934theuse}.

\begin{proposition}
\label{prop:mc-bounds}
Fix the candidate boundaries before generating the independent simulation samples. If the total tail probability is $\eta$ and the two one-sided bounds use tail probability $\eta/2$, then, with probability at least $1-\eta$ over these samples, both true least-favorable error probabilities are no greater than $U_K$ and $U_H$, respectively.
\end{proposition}

The proposition concerns one independent simulation after the candidate boundaries have been fixed. If another independent simulation is conducted after the first result, the tail probabilities must be allocated across the repeated simulations. When the exact error probabilities can be calculated, those probabilities should be reported.

\section{Proofs and Additional Details}
\label{app:proofs}

\subsection{Proof of Theorem~\ref{thm:least-favorable}}

Consider two populations with $m_1<m_2$ deviations. Construct the second population from the first by changing $m_2-m_1$ nondeviating items into deviations. Apply the same random permutation to both populations. The resulting cumulative counts satisfy
\begin{equation}
 S_t^{(m_1)}\leq S_t^{(m_2)}
 \quad\text{for every }t.
 \label{eq:path-order}
\end{equation}

Under boundaries satisfying $a_t<b_t$, selecting $K$ is an increasing event in the cumulative-count path. Suppose that the path for $m_1$ first selects $K$ at time $t$. The path for $m_2$ cannot have selected $H$ at an earlier time because $S_j^{(m_1)}\leq S_j^{(m_2)}$ and selection of $H$ requires $S_j\leq a_j$. At time $t$, we have $S_t^{(m_2)}\geq S_t^{(m_1)}\geq b_t$, so the path for $m_2$ has selected $K$ by time $t$. The same argument with the order reversed shows that selecting $H$ is a decreasing event. The terminal decision is also monotone because it selects $K$ exactly when $m>m_r$. The decision probabilities therefore satisfy
\begin{equation}
 \Pr_{m_1}(\delta=K)\leq\Pr_{m_2}(\delta=K)
 \quad\text{and}\quad
 \Pr_{m_1}(\delta=H)\geq\Pr_{m_2}(\delta=H).
 \label{eq:decision-monotonicity}
\end{equation}
Taking the largest count in $H$ and the smallest count in $K$ gives \eqref{eq:least-favorable-errors}.

\subsection{Boundaries at One Decision Time}
\label{app:boundary-property}

Fix the boundaries before decision time $t$. Let $\mathcal F_t$ be the set of integer pairs $(a,b)$ with $-1\leq a<b\leq t+1$ that satisfy the two cumulative inequalities in \eqref{eq:recursive-boundaries}. Define
\begin{equation}
 w_t(a,b)=b-a-1,
 \label{eq:continuation-state-count}
\end{equation}
which is the number of cumulative deviation counts for which the procedure continues at time $t$.

Let $a_t^*$ be the largest lower boundary that satisfies the second inequality in \eqref{eq:recursive-boundaries}, and let $b_t^*$ be the smallest upper boundary that satisfies the first inequality. Every feasible pair satisfies $a\leq a_t^*$ and $b\geq b_t^*$. If $a_t^*<b_t^*$, the pair $(a_t^*,b_t^*)$ therefore minimizes $w_t(a,b)$ over $\mathcal F_t$. If $a_t^*\geq b_t^*$, any pair $(c,c+1)$ with $b_t^*-1\leq c\leq a_t^*$ has no continuation state and attains the minimum value zero. The tie rule in Section~\ref{sec:procedure} selects one of these pairs.

This property concerns the current decision time after the earlier boundaries have been fixed. It does not state that the complete boundary sequence minimizes continuation probability or expected sample size.

\subsection{Expected Sample Size and Number of Interim Decisions}

Let $h_t^{(m)}$ and $k_t^{(m)}$ denote the probabilities of first selecting $H$ and $K$ at time $t$. Under complete inspection at $n$, the expected sample size is
\begin{equation}
 \bbE_m[\tau]=\sum_{t=1}^{n}t\left(h_t^{(m)}+k_t^{(m)}\right).
 \label{eq:expected-sample}
\end{equation}
Equivalently,
\begin{equation}
 \bbE_m[\tau]=\sum_{t=0}^{n-1}\Pr_m(\tau>t).
 \label{eq:tail-sum}
\end{equation}

If interim decisions are allowed at times $d_1<\cdots<d_J$, let $D$ be the number of these times reached. Its expectation is
\begin{equation}
 \bbE_m[D]=\sum_{j=1}^{J}\Pr_m(\tau\geq d_j).
 \label{eq:expected-decisions}
\end{equation}
The notebooks compute both quantities from the same continuation probabilities used in the boundary recursion.

\subsection{Computation Time}

At time $t$, the cumulative deviation count has at most $t+1$ states. The forward recursion therefore requires $O(n^2)$ arithmetic operations and $O(n)$ working memory for one deviation count. The boundary construction uses the two counts $m_H$ and $m_K$. Evaluation over every $m=0,\ldots,n$ requires additional calculations but is not needed to construct the boundaries.

\section{Additional Numerical Results}
\label{app:additional-results}

\subsection{Population Size, Tolerable Deviation Rate, and Error Limits}
\label{app:favorable-endpoint-sensitivity}

\begin{figure}[H]
\centering
\includegraphics[width=0.82\linewidth]{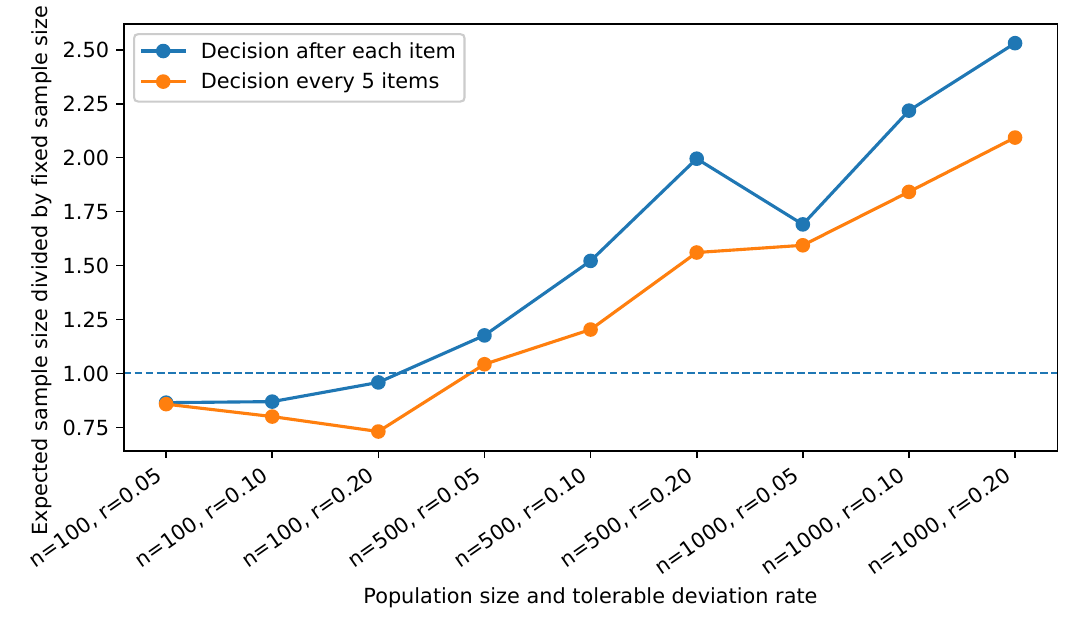}
\caption{Expected sample size relative to the exact fixed sample size at $m_H$.}
\label{fig:audit-inputs-favorable}
\end{figure}

\begin{figure}[H]
\centering
\includegraphics[width=0.82\linewidth]{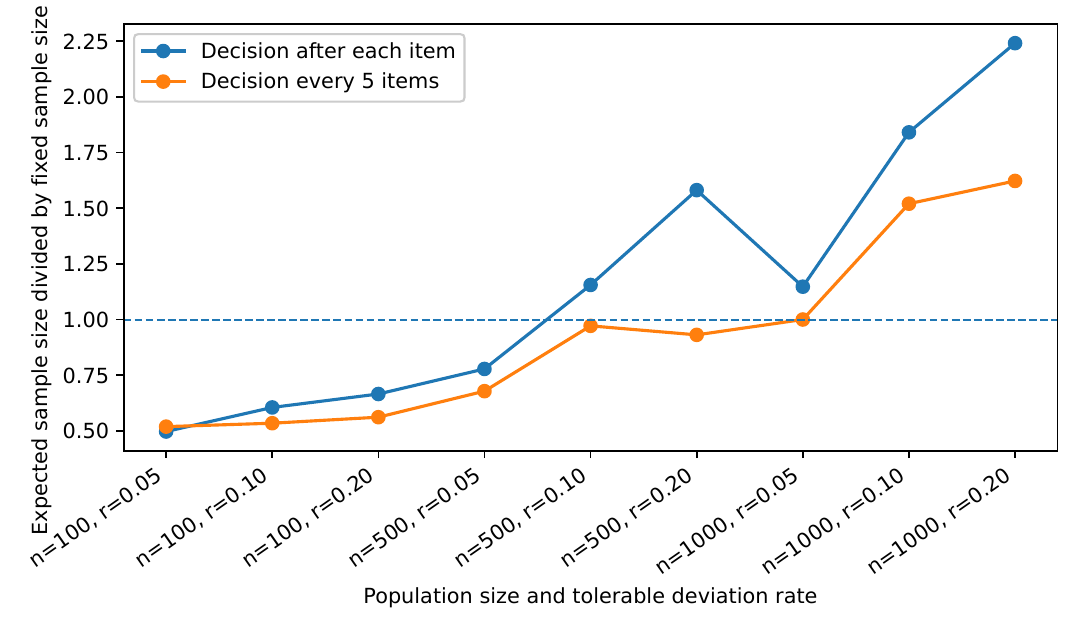}
\caption{Expected sample size relative to the exact fixed sample size at $m_K$.}
\label{fig:audit-inputs-unfavorable}
\end{figure}

The calculations in Table~\ref{tab:audit-inputs} use $\alpha=\beta=0.05$. Figures~\ref{fig:audit-inputs-favorable} and \ref{fig:audit-inputs-unfavorable} report the expected sample size divided by the exact fixed sample size at $m_H$ and $m_K$, respectively. A ratio above one means that the sequential procedure has a larger expected sample size at that population count.

The accompanying CSV file also reports exact calculations with $\alpha=\beta=0.10$. The larger error limits reduce the sample sizes for both the sequential and fixed-sample procedures. Which procedure has the smaller expected sample size depends on the population count and the times at which decisions are allowed.

Table~\ref{tab:favorable-endpoint-sensitivity} holds $n=100$, $r=0.20$, $m_K=21$, and $\alpha=\beta=0.05$ fixed and varies $m_H$. Decisions are permitted after each item.

\begin{table}[H]
\centering
\caption{Sensitivity to the endpoint of the favorable region}
\label{tab:favorable-endpoint-sensitivity}
\begin{tabular}{rrrrr}
\toprule
$m_H$ & Exact fixed sample & $\bbE_{m_H}[\tau]$ & $\bbE_{21}[\tau]$ & $\max_m \bbE_m[\tau]$ \\
\midrule
10 & 57 & 54.63 & 37.95 & 59.57 \\
15 & 82 & 78.28 & 61.02 & 78.28 \\
18 & 96 & 85.63 & 77.19 & 85.63 \\
\bottomrule
\end{tabular}
\end{table}

Moving $m_H$ closer to the tolerable count increases the expected sample size near both protected boundary counts in this example.

\subsection{Asymmetric Error Limits}

We set $n=200$, $r=0.10$, $m_H=16$, $m_r=20$, $m_K=21$, and $\alpha=0.05$. The value of $\beta$ is 0.05, 0.025, or 0.01.

\begin{figure}[H]
\centering
\includegraphics[width=0.80\linewidth]{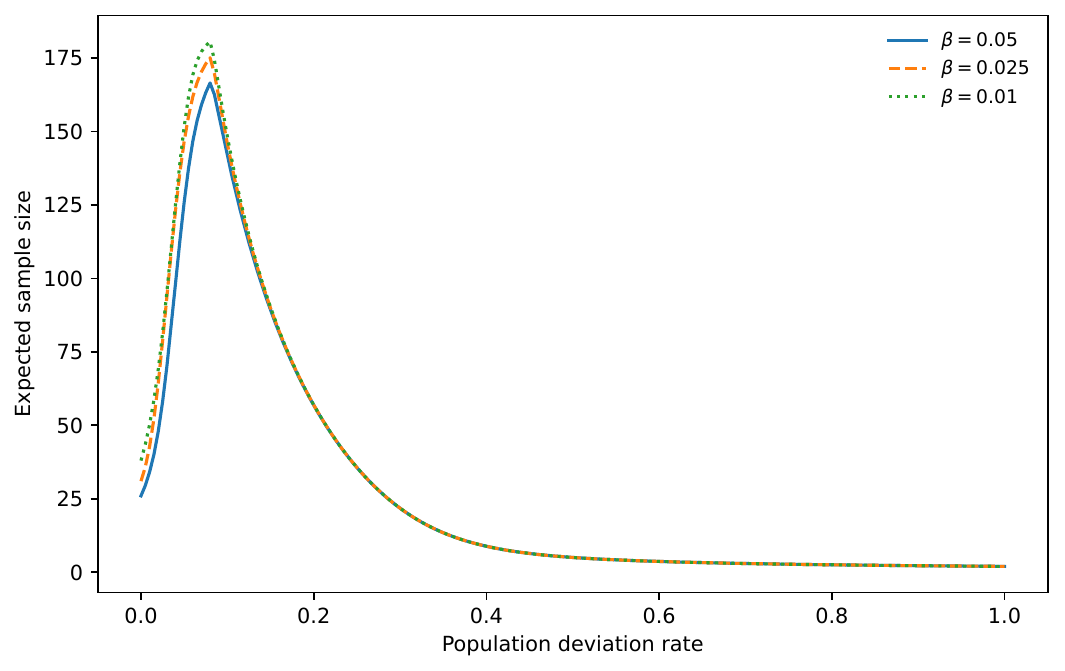}
\caption{Expected sample size under different limits on the risk of assessing control risk too low.}
\label{fig:asymmetric-errors}
\end{figure}

At $m=21$, the probability of selecting $H$ equals 0.05, 0.025, and 0.01 for the three values of $\beta$. The expected sample size at $m=16$ increases from 166.49 to 180.51 as $\beta$ decreases from 0.05 to 0.01. The stricter limit mainly delays selection of $H$.

\subsection{Fixed Maximum Sample Size}

For the basic setting, we allow interim decisions after every 10 items and require a final decision no later than $T$. We set the interim limits to $\alpha_I=\beta_I=0.04$, leaving part of each total limit for the final decision. The smallest feasible maximum sample size is $T=89$, and the terminal rule selects $H$ when $S_{89}\leq15$.

The resulting probabilities are 0.03991 for selecting $K$ at $m=15$ and 0.04802 for selecting $H$ at $m=21$. The expected sample size is 75.62 at $m=15$ and 59.80 at $m=21$. The probabilities of reaching the maximum sample size are 0.7640 and 0.1621, respectively.

\begin{figure}[H]
\centering
\includegraphics[width=0.78\linewidth]{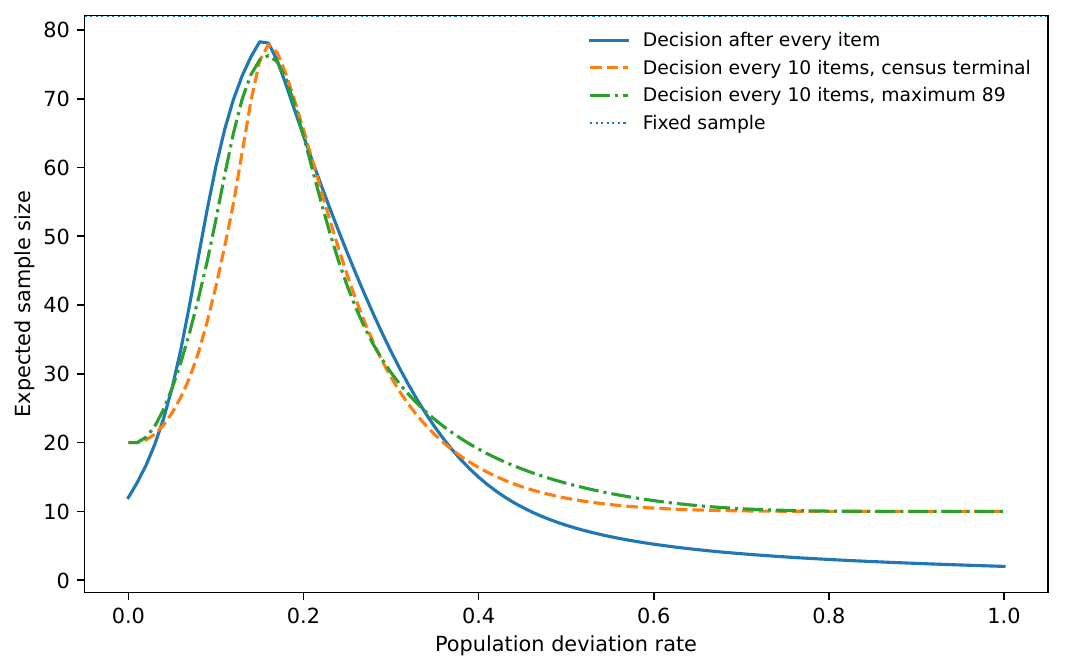}
\caption{Expected sample size for the procedure with maximum sample size 89.}
\label{fig:fixed-maximum}
\end{figure}

\subsection{Numerical Results for the Monte Carlo Approximation}

We construct candidate boundaries with 20,000 random permutations at $m_H$ and $m_K$ and repeat the calculation for 30 seeds. When the internal limits equal the target value 0.05, exact reevaluation exceeds 0.05 in 60.0 percent of repetitions for selecting $K$ at $m_H$ and in 56.7 percent for selecting $H$ at $m_K$. The largest exact values are 0.05390 and 0.05347.

When the internal limits equal 0.046, none of the 30 exact reevaluations exceeds 0.05. The largest exact values are 0.04945 and 0.04959. The 30 repetitions do not establish that an internal limit of 0.046 is sufficient in every setting.

\begin{figure}[H]
\centering
\includegraphics[width=0.82\linewidth]{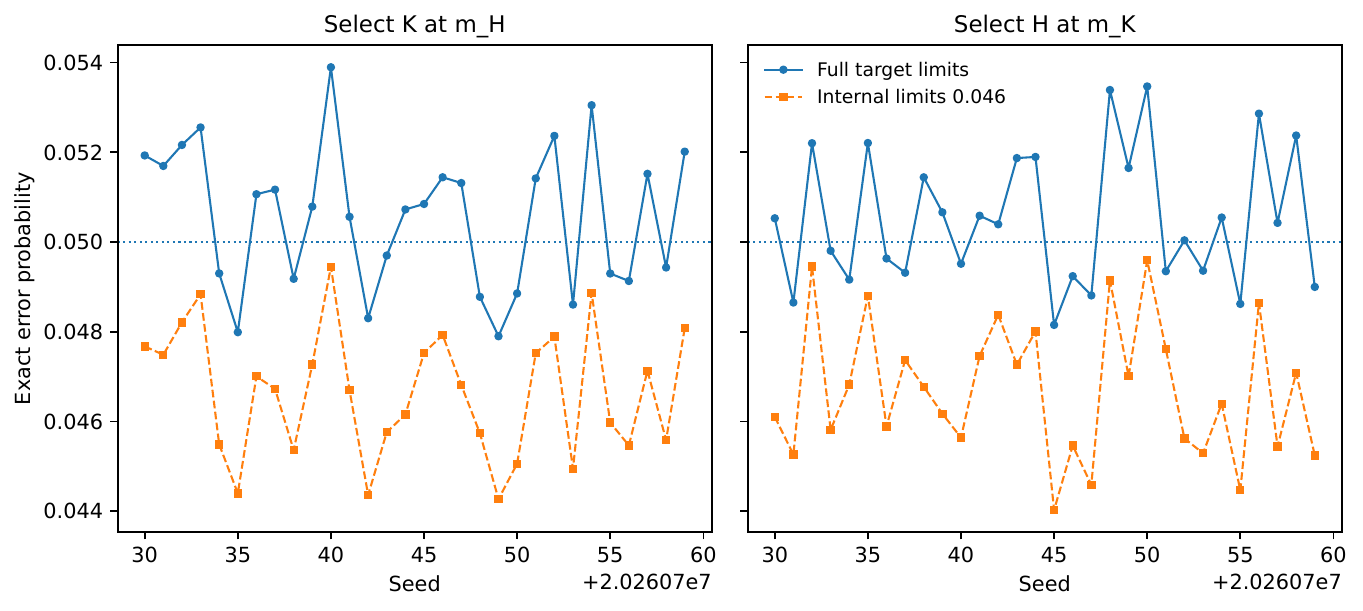}
\caption{Exact error probabilities of procedures obtained from 20,000 Monte Carlo paths.}
\label{fig:monte-carlo}
\end{figure}

For a separate candidate constructed with 100,000 paths at each of $m_H$ and $m_K$ and internal limits of 0.046, we generate 200,000 independent paths in each direction and set the total tail probability to 0.01. The one-sided upper bounds are 0.04790 and 0.04738. Exact calculation for the fixed candidate gives 0.04613 and 0.04634.

\subsection{Complete Examination of Selected Items}

\begin{figure}[H]
\centering
\includegraphics[width=0.80\linewidth]{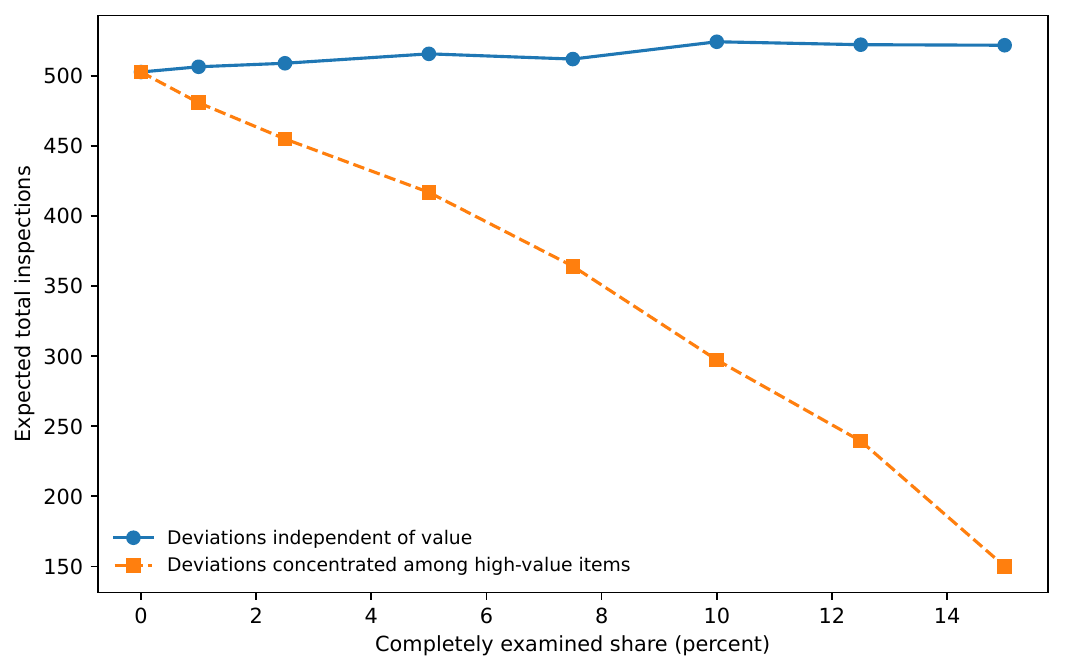}
\caption{Expected total inspections after complete examination of high-value items.}
\label{fig:selected-items}
\end{figure}

We use two synthetic populations of 1,000 items to illustrate complete examination of high-value items followed by random sampling of the residual population. Both populations contain 130 deviations. In the first population, deviations are assigned independently of value. In the second, deviations are more concentrated among high-value items. Figure~\ref{fig:selected-items} reports the expected total number of inspected items for selected shares from zero to 15 percent.

When deviations are independent of value, examining more high-value items does not consistently reduce the total number inspected. When deviations are concentrated among those items, the observed deviations change the residual thresholds and can reduce the remaining number of inspections. The calculation is illustrative because the simulation specifies both the rule for selecting items and the assignment of deviations.

\subsection{Illustration Using CMS CERT Claims}
\label{app:cert}

The CMS Comprehensive Error Rate Testing program draws a stratified random sample of Medicare fee-for-service claims and uses an independent medical review contractor to determine whether each claim was paid properly under coverage, coding, and billing rules \citep{CMS2026medicarefeeforservice,CMS2026comprehensiveerror}. The 2025 public file contains claim-line review results. We aggregate the lines to the claim level and define an illustrative deviation when at least one line has \texttt{Review Decision} equal to \texttt{Disagree}.

The resulting data contain 37,581 claims. We analyze Part B and durable medical equipment Medicare Administrative Contractor claims separately. For Part A, one subset excludes claims under the inpatient prospective payment system, and the other includes them. Table~\ref{tab:cert-summary} reports the fixed number of claims and the proportion of claims classified as deviations in each subset.

\begin{table}[H]
\centering
\small
\caption{CMS CERT claim-level populations}
\label{tab:cert-summary}
\begin{tabular}{lrrr}
\toprule
Claim type & Claims & Deviations & Deviation rate \\
\midrule
Part B & 11,530 & 2,312 & 0.2005 \\
DME MAC & 8,650 & 1,703 & 0.1969 \\
Part A, non-IPPS & 8,651 & 1,298 & 0.1500 \\
Part A, IPPS & 8,750 & 1,514 & 0.1730 \\
\bottomrule
\end{tabular}
\end{table}

The released claims were drawn through the CMS stratified sampling design. We condition on the released claims and treat each subset defined by claim type as a fixed finite population. Under a uniformly random inspection order, the calculation for a subset depends on its number of claims and number of claim-level disagreements. We do not use the CMS sampling weights and do not estimate the Medicare improper-payment rate.

For illustration, we set $r=0.18$, $\theta_H=0.01$, and $\alpha=\beta=0.05$. The value 0.18 is close to the overall claim-level disagreement rate in the released file and is not a CMS audit threshold. The unfavorable region begins at the first claim count above the tolerable count. We compare decisions after each item with decisions after every 100 items.

\begin{table}[H]
\centering
\small
\caption{Exact results for the CMS CERT claim-type subsets}
\label{tab:cert-results}
\begin{tabular}{llrrr}
\toprule
Claim type & Decision interval & Expected sample & $\Pr(H)$ & $\Pr(K)$ \\
\midrule
Part B & 1 item & 4,698.40 & 0.0320 & 0.9680 \\
Part B & 100 items & 4,201.89 & 0.0130 & 0.9870 \\
DME MAC & 1 item & 4,508.39 & 0.0347 & 0.9653 \\
DME MAC & 100 items & 4,055.46 & 0.0166 & 0.9834 \\
Part A, non-IPPS & 1 item & 4,192.27 & 0.9637 & 0.0363 \\
Part A, non-IPPS & 100 items & 3,467.37 & 0.9881 & 0.0119 \\
Part A, IPPS & 1 item & 7,510.14 & 0.9027 & 0.0973 \\
Part A, IPPS & 100 items & 7,341.92 & 0.9354 & 0.0646 \\
\bottomrule
\end{tabular}
\end{table}

\begin{figure}[H]
\centering
\includegraphics[width=0.78\linewidth]{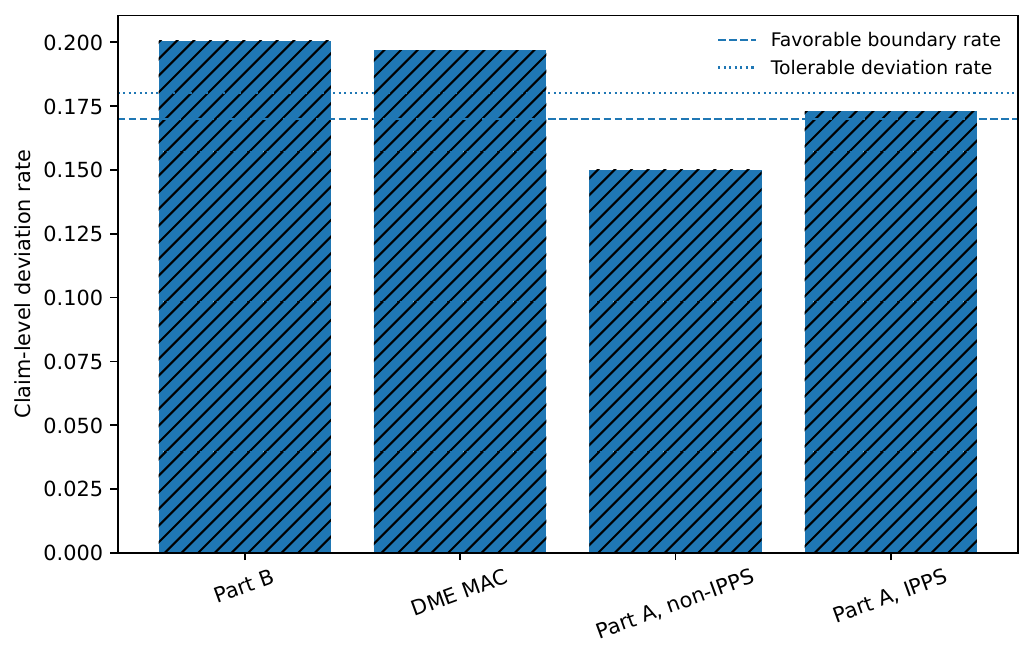}
\caption{Claim-level deviation rates in the four CMS CERT subsets.}
\label{fig:cert-rates}
\end{figure}

\begin{figure}[H]
\centering
\includegraphics[width=0.82\linewidth]{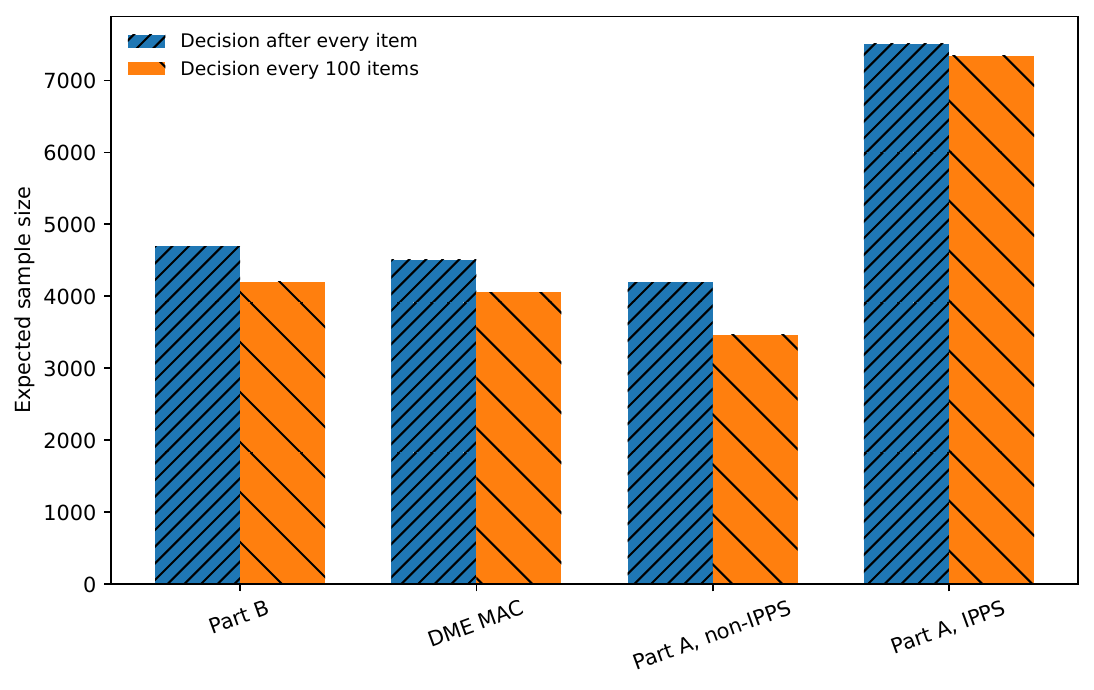}
\caption{Expected sample size in the four CMS CERT subsets.}
\label{fig:cert-sample}
\end{figure}

The released CERT claims were selected under the CMS stratified design, and the public file does not report the order in which reviewers examined claims. The calculation applies a random inspection order within each subset defined by claim type. It does not evaluate the estimator used by CMS, reproduce the order in which CMS reviewed the claims, or establish audit-hour savings.

\end{document}